# On the analysis of the tin-inside-H₃S Mössbauer experiment

Ruslan Prozorov and Sergey L. Bud'ko

*Ames Laboratory and Department of Physics & Astronomy, Iowa State University, Ames, Iowa 50011, USA*

13 April 2022

## Abstract

A simple analysis is presented of the particular experiment used to prove the bulk nature of very-high-$T_c$ superconductivity in $H_3S$ compound under ultra-high pressure. In the experiment, an internal magnetic field was sensed by the synchrotron Mössbauer spectroscopy in tin placed inside the $H_2S$ sample. The experiment showed peculiar anisotropy with respect to the direction of the applied field at first sight. By considering actual experimental geometries and parameters of the experiment, we show that this particular observation is consistent with the expectations for a regular type-II superconductor with Meissner expulsion and pinning.

## Introduction

Since the discovery of superconductivity at ~200 K in $H_3S$ (formed from $H_2S$) [1], similar or higher transition temperatures, $T_c$, have been reported in various hydrogen-rich compounds under ultra-high pressures [2]. Considering quite restrictive conditions for the sample in a diamond anvil pressure cell (DAC), every "simple" experiment becomes challenging. Additionally, due to tiny samples and massive surroundings, the signal to background ratio is also very small. So far, the crystal structures of the superconducting phases were determined by X-ray diffraction, and one "must-have" feature of superconductivity, the drop in the electrical resistance, was experimentally shown [1]. Of course, electrical transport alone cannot be considered proof of superconductivity. The second "must-have" property is the diamagnetic screening of the magnetic field. Note that we do not use the term "Meissner expulsion" because some best strong superconductors exhibit almost no expulsion, and there are various reasons for that [3]. Although one can imagine a hypothetical "perfect metal" that would screen the applied magnetic field, in practice, the substantial screening of DC magnetic fields only exists in superconductors.

Usually, the analysis of zero-field cooled (ZFC) and field cooled (FC) magnetic susceptibility measurements of samples at ambient pressure would be used to provide such evidence. Indeed, in the discovery publication [1], the magnetic field screening in the superconducting state in ZFC measurements was clearly observed. However, these measurements impose additional restrictions on the size, design, and materials of the diamond anvil cell (DAC) and the shape and continuity of the superconducting sample [1,4].

An alternative approach to establishing the magnetic footprint of superconductivity under very high pressures was implemented in Ref.[5], where a thin $^{119}$Sn foil placed inside the sample in DAC was used as a sensor of the internal magnetic field monitored by nuclear resonance scattering of synchrotron radiation. Although the addition of Sn foil inside the sample made the DAC preparation more complex, these measurements placed less (or no) restrictions on DAC design, required no contacts or coils, and could be realized using dedicated beamlines in several synchrotron facilities.

Following the publication, the interpretation of the results of Ref.[5] was criticized [6-9] with the suggested outcome could be inconsistent with superconductivity. Since the question of superconductivity above 200 K is of significant interest to the community, in this work, we re-examine the experimental data of Ref.[5] considering actual experimental geometry. Our analysis is consistent with the behavior of a type-II superconductor with moderate pinning in the nuclear resonance scattering experiments [5].

### London-Meissner state

Let us analyze the results shown in Fig.4 of Ref.[5]. Considering the relatively small sizes of the components involved, it is not apparent what degree of screening and spatial field distributions are expected in these experiments. The substantial difference in the measured internal magnetic field for two orientations may seem puzzling, but in fact, it is expected. We use published results to estimate relevant parameters that, indeed, are consistent with a generic bulk type-II superconductor with moderate pinning. Two phenomena are relevant to the problem at hand, Meissner – London screening [10] and Bean critical state when Abrikosov vortices penetrate the sample in a non-uniform manner [11]. Due to the small sizes and finite shapes involved, London equations need to be solved numerically for this exact geometry and finite London penetration depth.

We used the commercial software environment COMSOL 5.6 with AC/DC module to perform an adaptive finite element study [12]. Technical aspects of the three-dimensional (3D) modeling and methods of solving London equations in arbitrarily shaped superconductors, including cylinders, are given elsewhere [13,14]. A comprehensive study of the Meissner-London state in finite cylinders is described in Ref.[13]. We note, however, that in that paper, the modeling was done on a 2D cross-section and used axial rotational symmetry for the 3D analysis. In such a case, the adaptive mesh is constructed on a two-dimensional cross-section of the assembly, significantly reducing computing requirements for calculations, hence allowing for a much higher spatial resolution. Here we use a full 3D adaptive element scheme because the magnetic field parallel to the disc breaks rotational symmetry. We verified that the axial case presented here is fully consistent with the results of Ref.[13]

The geometry of the simulation is shown in Fig.1. The superconductor is a disk, 30 μm in diameter and 5 μm thick. Inside, it contains a cylindrical cavity filled with tin, which is not superconducting at the temperatures of interest. The tin cavity is 20 μm in diameter and 2.5 μm thick. This structure is fixed in space with a short dimension parallel to the vertical z-axis so that the disc itself is in the xy-plane. An external magnetic field can be applied either perpendicular to the disc or parallel. We solved the London equations and determined the magnetic induction distribution inside and outside the sample assembly. The convergence of the solutions was verified by changing the parameters and types of the adaptive mesh elements.

Figure 2 shows the distribution of the amplitude of the magnetic induction in 2D cross-sections of the sample assembly for two orientations of the applied magnetic field of 6500 Oe (from the actual experiments of Ref.[5]) and for two values of London penetration depth, λ=0.2 μm and λ=1.3 μm. The first value is chosen as a typical representative number for a generic superconductor corresponding to a lower critical field, $B_{c1} \approx 100\,\text{G}$ assuming a moderate Ginsburg-Landau parameter, $\kappa = \lambda/\xi = 6$, also quite typical and estimated earlier for systems like ours. The second value of λ=1.3 μm was chosen to obtain a magnetic field of 0.27 T inside the tin–filled cavity in a parallel orientation, Fig.2(d). However, this value corresponds to $B_{c1} \approx 2\,\text{G}$, which is unreasonable. More importantly, though, in this case, the magnetic field in tin space in the perpendicular orientation is far from zero, as can be seen in Fig.2(b) and is quantitatively explained in Fig.3.

Figure 3 shows magnetic induction profiles across the sample for perpendicular (top) and parallel (bottom) orientation in the direction perpendicular to the applied magnetic field. The profiles are shown for the three values of London penetration depth that, in addition to those already discussed, include λ=20 nm, corresponding to $B_{c1} = 1\,\text{T}$, which was suggested in some previous ultra-high-pressure works [1]. At this value of λ, a magnetic field is well shielded in both orientations. However, the corresponding $B_{c1}$ exceeds the external magnetic field applied in the analyzed experiment ($B_{ext} = 0.68\,\text{T}$ and 0.65 T for perpendicular and parallel orientations, respectively), so it is impossible.

To analyze the plausible range of λ from the Meissner-London state point of view, we calculated the magnetic induction averaged over the volume of the Sn pocket for a range of λ values for two orientations of a magnetic field. The result is shown in Figure 4. After a flat region up to λ approximately 0.3 μm, the average field starts to increase, expectedly faster for the parallel orientation due to small thickness. Note, the demagnetization correction is automatically taken into account in our calculations, see Fig.3 (top). The green path in Fig.4 shows that in order to reproduce the magnetic field measured in a parallel orientation, $B_0 \approx 0.27\,\text{T}$, we need a London penetration depth of λ=1.3 μm. However, at this λ, a smaller but still well measurable field $B_0 \approx 0.06\,\text{T}$ is predicted in the perpendicular orientation, but that was not observed.

This part concludes that we cannot fully explain the results based solely on the London-Meissner physics alone. Let us turn to a magnetic flux penetration in the form of vortices that occurs when a magnetic field exceeds the first critical field $B_{c1}$.

### The Bean critical state

When a magnetic field on the sample surface exceeds the lower critical field, it starts entering the sample in the form of Abrikosov vortices. The pinning forces due to imperfections, defects, and impurities prevent vortices from free motion and produce an opposite force that balances the Lorentz force that pushes the vortices inside. The simplest model describing this process was suggested by Bean [11]. It considers constant, field-independent critical current density, $\mathbf{J_c}$. According to the Maxwell equation, $\nabla \times \mathbf{H} = \mathbf{J_c}$, it results in a constant gradient of the magnetic induction inside the sample, roughly $dB/dx = \mu_0 J_c$. Even for isotropic critical current density, the field at which vortices reach the center, known as the field of full penetration, $B^*$, depends on the distance, $a$, from the edge to the center in a direction perpendicular to the direction of the magnetic field, $B^* = \mu_0 J_c a$. Since this distance is different for the two orientations, field $B^*$ is also different. While we do not know how far the magnetic field reaches inside in the perpendicular orientation, we can use the parallel orientation to estimate $J_c$. The magnetic field in the center is about $B_0 = 0.25\,\text{T}$, and the applied field is $B_{ext} = 0.65\,\text{T}$. If we neglect the London-Meissner edge drop, $B_{c1}$, we obtain, $B^* = 0.65 - 0.25 = 0.4\,\text{T}$. Therefore, the critical current density, $J_c = B^*/(\mu_0 c) \approx 12.7\,\text{MA/cm}^2$, where $c = 2.5\,\mu\text{m}$ is the half-thickness of the superconducting disc. This is a substantial current density because the values of the order of $\sim 1\,\text{MA/cm}^2$ are already considered large and suitable for practical applications. Ten times that is, in principle, possible but highly unlikely, especially at such high temperatures. In a more realistic scenario, we do have a surface decrease of the magnetic induction due to Meissner-London shielding. The uncertainty can be reduced by noting that flux penetrating in the perpendicular orientation does not reach the volume exposed to the γ-rays beam. With a γ-beam diameter of 15 μm, the distance from the edge to the beam in the radial direction is (30-15)/2=7.5 μm. The distance from the edge to the vortex front in the critical state is $x = B^*/\mu_0 J_c$. Therefore, the smallest critical current allowed in this model corresponds to $x = 7.5\,\mu\text{m}$. It is trivial to obtain, $B_{c1} = B_{ext} - B_0/(1 - c/x)$, indeed, only valid for $x > c$. For $x = 7.5\,\mu\text{m}$ and $c = 2.5\,\mu\text{m}$ it gives a reasonable value of $B_{c1} \approx 0.28\,\text{T}$, and the corresponding critical current density is, $J_c \approx 4\,\text{MA/cm}^2$, which is the reasonable and relatively small current density.

We note that our analysis is not related to the question of the conventional or unconventional mechanism of superconductivity, only to its type in terms of the behavior in a magnetic field. For example, Eliashberg theory analysis suggests significant strong-coupling and retardation effects [15].

## Conclusions

Based on the analysis of London-Meissner and Bean critical state, we conclude that the experiment in which the average magnetic field was measured in the interior cavity of a $H_2S$ superconductor in two orientations by resonant Mössbauer is consistent with the behavior expected of a type-II superconductor with the lower critical field, $0.3 \lesssim B_{c1} \lesssim 0.6\,\mathrm{T}$ and critical current density larger than $J_c \approx 4\,\mathrm{MA/cm^2}$. Of course, these estimates are approximate, and more detailed magnetic measurements in different experimental conditions are needed to fully understand this fascinating, almost room-temperature superconductivity.

## Acknowledgments:

We acknowledge discussions with various parties involved in the ultra-high-pressure superconductivity research. RP especially thanks J. E. Hirsch for numerous comments and insightful discussions. This work was supported by the U.S. Department of Energy (DOE), Office of Science, Basic Energy Sciences, Materials Science and Engineering Division. Ames Laboratory is operated for the U.S. DOE by Iowa State University under contract DE-AC02-07CH11358.

Figures:

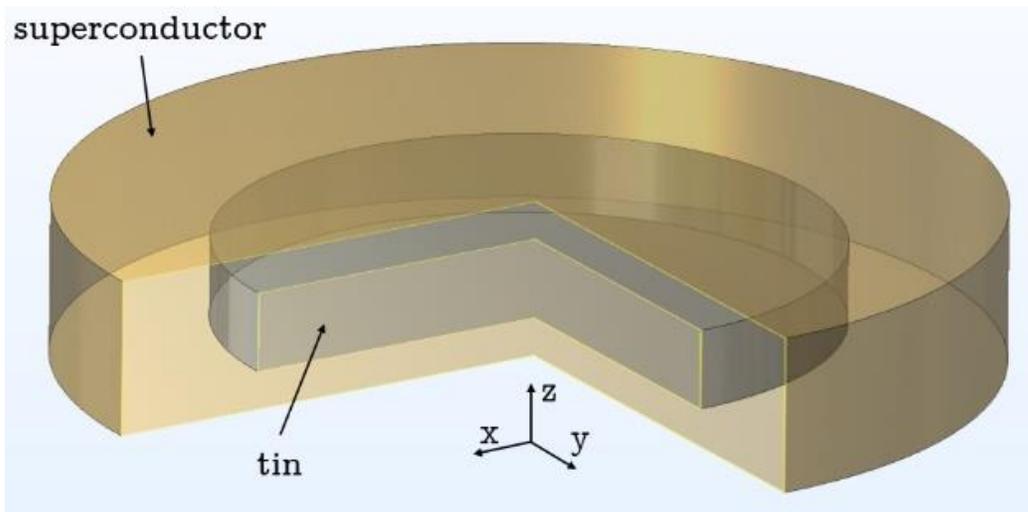

Figure 1: Geometry of the numerical analysis. Superconducting disc, 30 µm in diameter and 5 µm thick, contains a disc – shaped enclosure, 20 µm in diameter and 2.5 µm thick. In the experiment, this enclosure is filled with non-superconducting (at temperatures of interest) tin.

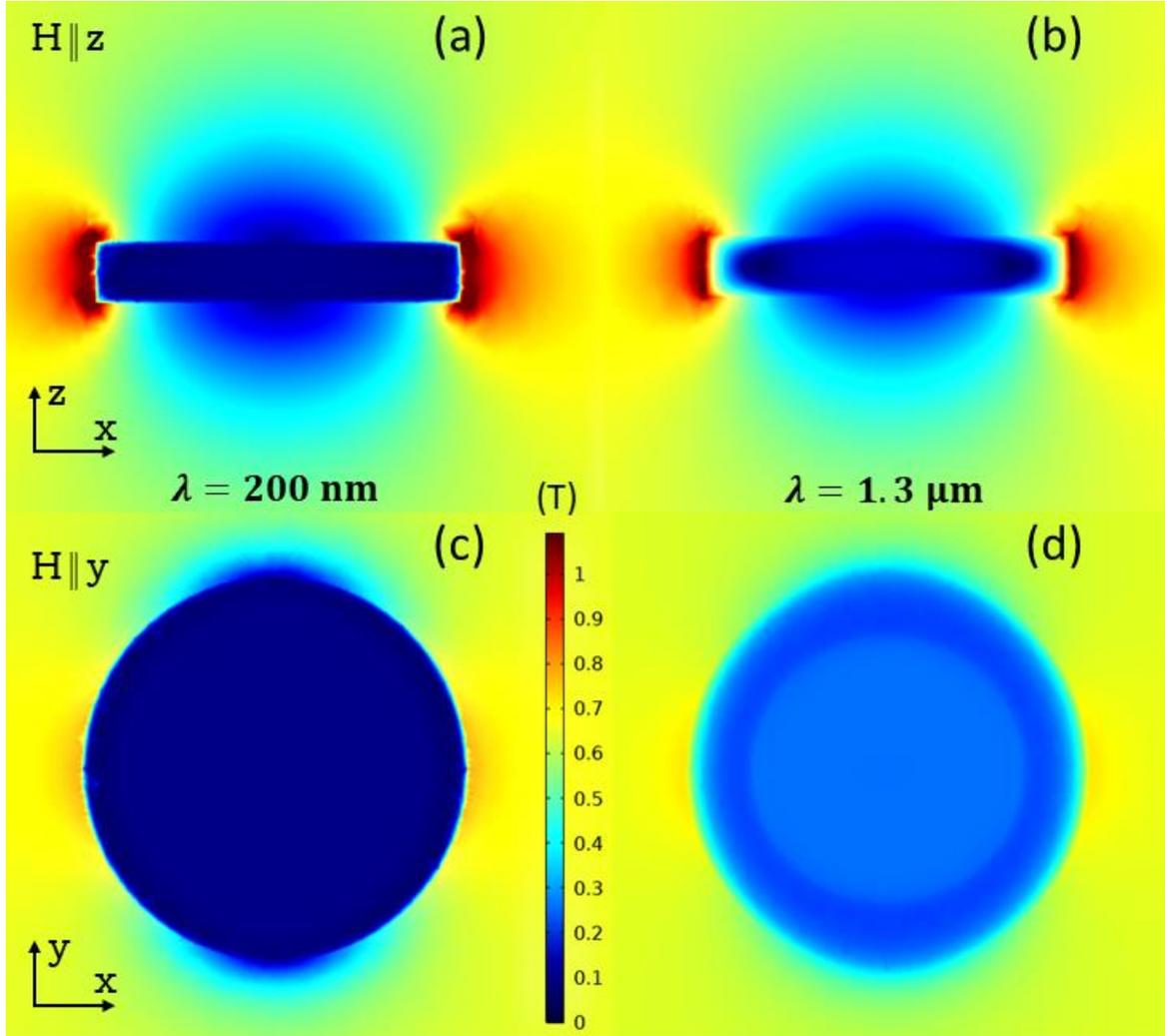

Figure 2: Two dimensional maps of the magnetic induction in principal cross-sections for two values of London penetration depth shown with a common color scale. A magnetic field of 0.65 T is applied. Panels (a) and (b) show the xz-plane cross-section for the magnetic field oriented perpendicular to the disc, along the z-axis. Panels (c) and (d) show the xy-plane cross-section with a magnetic field applied along the y-axis. Panels (a) and (c) correspond to London penetration depth, λ=200 nm, whereas panels (b) and (d) show the results for λ=1300 nm. We explain the selection of these particular values in the text. As can be seen from a much lighter color in panel (d), the magnetic induction inside the disc is far from zero, in fact it is equal to 0.27 T, matching the experimental observations. However, the field is also quite large in the other orientation, panel (b), which is not what we observed. The explanation requires another mode of flux penetration exceeding lower critical field, $B_{c1}$, see text for discussion.

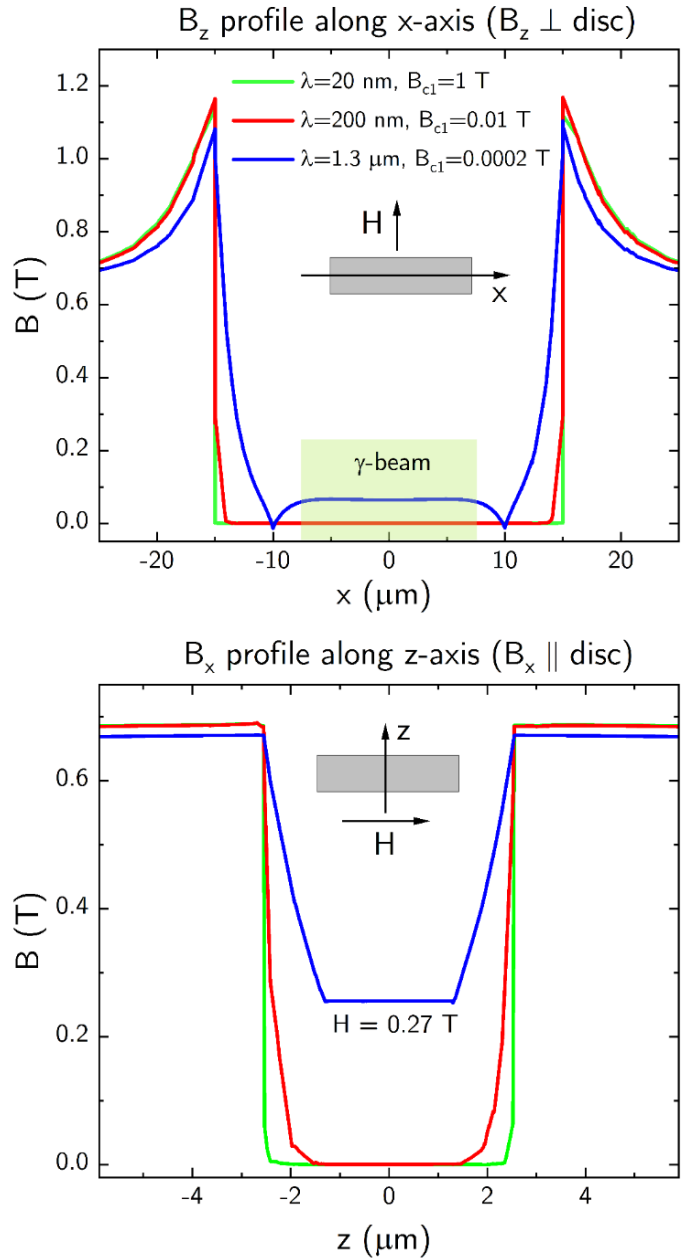

Figure 3. Magnetic induction profiles across the sample in case of an applied magnetic field (shown schematically) parallel to the z-axis and perpendicular to the disc plane (top), and for a magnetic field parallel to the y-axis and parallel to the disc plane. The profiles are shown for three values of London penetration depth that, in addition to already discussed values, includes the value of λ=1.3 μm, corresponding to $B_{c1} \approx 2\,\text{G}$.

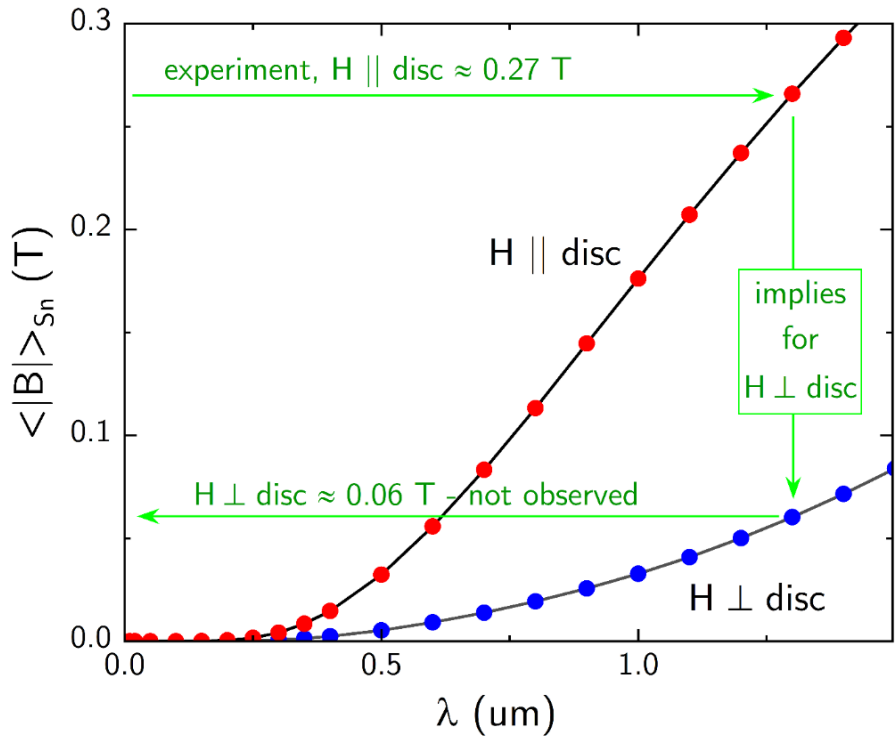

Figure 4: Average magnetic induction in Sn volume inside the superconducting sample. The green path shows that in order to reproduce the observed value of 0.27 T in a parallel orientation we need $\lambda$=1.3 µm and that would produce smaller, but still well measurable field of 0.06 T in the perpendicular orientation, but that was not observed.

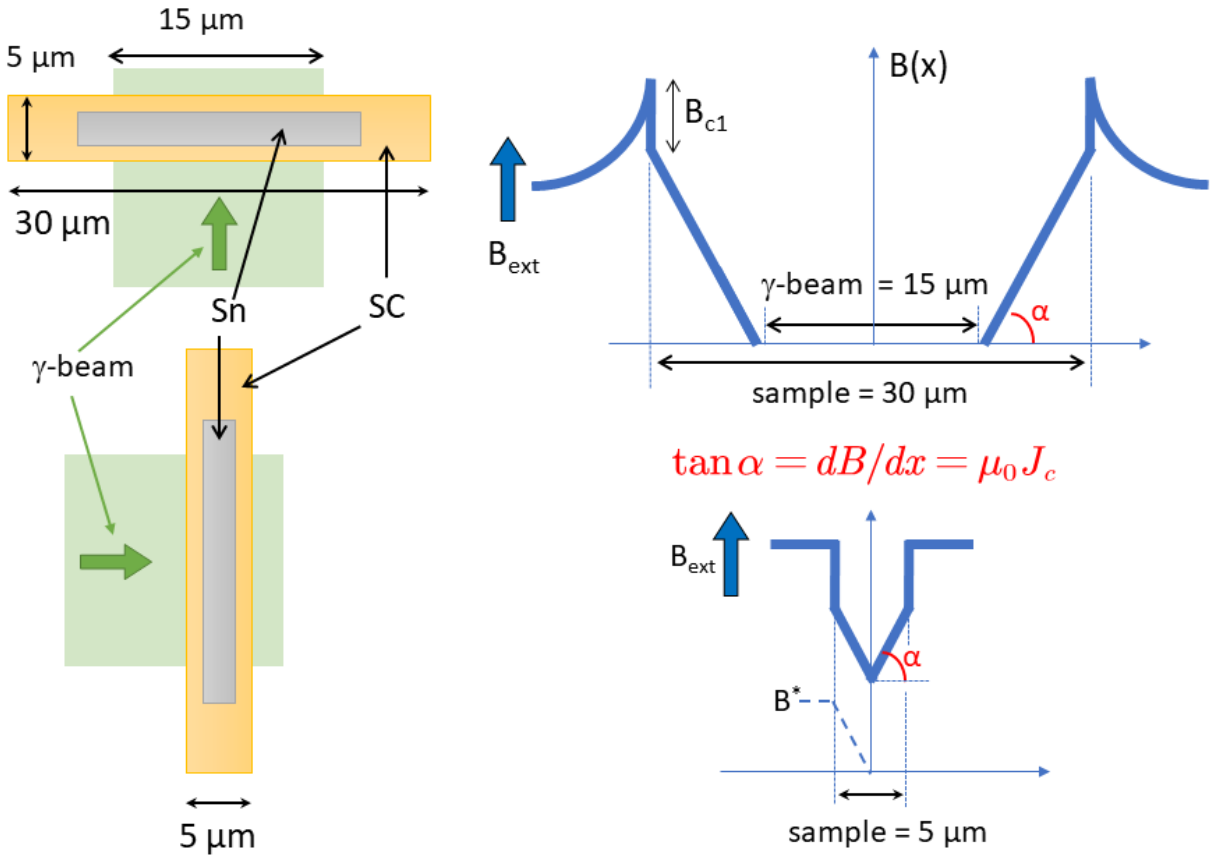

Figure 5: Left column shows the experimental arrangement. Here, to keep magnetic field vertical for the sketches in the right column, the sample assembly shown rotated 90 degrees. The top panel on the right shows flux penetration in the case of magnetic field perpendicular to the discs, the bottom shows the parallel arrangement. After the edge drop of magnetic induction by the value of $B_{c1}$ withing the London penetration depth layer of width λ (neglected in the sketch), linear profile of $B(x)$ is formed. The tangent of the slope is directly related to the critical current density, $\tan\alpha = dB/dx = \mu_0 J_c$
.